\documentclass[12pt]{article}

\usepackage{amsmath,amssymb,amsfonts,amsthm,hyperref,bbm,latexsym,epsfig}
\usepackage{graphicx,graphics,multirow}

\newcommand{\bs}{\begin{subequations}}
\newcommand{\es}{\end{subequations}}
\newcommand{\be}{\begin{equation}}
\newcommand{\ee}{\end{equation}}
\newcommand{\ba}{\begin{eqnarray}}
\newcommand{\ea}{\end{eqnarray}}
\newcommand{\no}{\nonumber \\}

\newcommand{\diag}{\mbox{diag}}
\newcommand{\mnu}{\mathcal{M}_\nu}

\allowdisplaybreaks

\begin{document}

\title{
\normalsize \hfill UWThPh-2016-27
\\[.5mm]
\normalsize \hfill CFTP/16-014
\\[4mm]
\LARGE Charged-lepton decays from soft flavour violation}

\author{
E.\ H.\ Aeikens,$^{(1)}$\thanks{E-mail: \tt elke.aeikens@univie.ac.at}
\addtocounter{footnote}{2}
W.~Grimus,$^{(1)}$\thanks{E-mail: \tt walter.grimus@univie.ac.at}
\ and
L.~Lavoura$^{(2)}$\thanks{E-mail: \tt balio@cftp.tecnico.ulisboa.pt}
\\*[3mm]
$^{(1)} \! $
\small University of Vienna, Faculty of Physics,
\small Boltzmanngasse 5, A--1090 Wien, Austria
\\[2mm]
$^{(2)} \! $
\small CFTP, Instituto Superior T\'ecnico, Universidade de Lisboa, Portugal
\\*[2mm]
}

\date{24 February 2017}

\maketitle

\begin{abstract}
We consider a two-Higgs-doublet extension of the Standard Model,
with three right-handed neutrino singlets and the seesaw mechanism,
wherein all the Yukawa-coupling matrices are lepton flavour-diagonal
and lepton flavour violation is \emph{soft},
originating solely in the non-flavour-diagonal Majorana mass matrix
of the right-handed neutrinos.
We consider the limit $m_R \to \infty$ of this model,
where $m_R$ is the seesaw scale.
We demonstrate that there is a region in parameter space
where the branching ratios of all five
charged-lepton decays $\ell_1^- \to \ell_2^- \ell_3^+ \ell_3^-$
are close to their experimental upper bounds,
while the radiative decays $\ell_1^- \to \ell_2^- \gamma$ are invisible
because their branching ratios are suppressed
by $m_R^{-4}$.
We also consider the anomalous magnetic moment of the muon
and show that in our model the contributions from the extra scalars,
both charged and neutral,
can remove the discrepancy between its experimental and theoretical values.  
\end{abstract}

\newpage

\section{Introduction}

In this paper we resume
an old idea of two of us~\cite{soft}:
in a multi-Higgs-doublet model
furnished with three right-handed neutrino singlets
and the seesaw mechanism~\cite{seesaw},
lepton flavour may be conserved
in the Yukawa couplings of all the Higgs doublets
and violated solely in the Majorana mass terms
of the right-handed neutrinos $\nu_{\ell R}$
($\ell=e,\mu,\tau$),
\textit{viz.}\ in
\be \label{MR}
\mathcal{L}_{\nu_R\, \mathrm{mass}} =
- \frac{1}{2}\, \sum_{\ell_1, \ell_2} \bar \nu_{\ell_1 R}
\left( M_R \right)_{\ell_1 \ell_2} C\, \bar \nu_{\ell_2 R}^T
+ {\rm H.c.},
\ee
where $C$ is the charge-conjugation matrix in Dirac space
and $M_R$ is a non-singular symmetric matrix in flavour space.
Since $\mathcal{L}_{\nu_R\, \mathrm{mass}}$ has dimension three,
the violation of the individual lepton flavour numbers $L_\ell$
and of the total lepton number $L = L_e + L_\mu + L_\tau$ is \emph{soft}.
Thus,
in our framework $\mathcal{L}_{\nu_R\, \mathrm{mass}}$ is responsible for 
\begin{enumerate}
\item the smallness of the light-neutrino masses,
\item lepton mixing,
\item violation of $L$, and
\item violation of $L_e$, $L_\mu$, and $L_\tau$.
\end{enumerate}

In this context,
lepton flavour-violating processes
were explicitly investigated at one-loop order in ref.~\cite{GL2002}
and the following property of our framework was discovered.
Let $m_R$ denote the seesaw scale---the scale of the square roots
of the eigenvalues of $M_R M_R^\ast$---and $n$ denote
the number of Higgs doublets; 
it was found in ref.~\cite{GL2002} that
\begin{enumerate}
\renewcommand{\labelenumi}{\roman{enumi}.}
\item
the amplitudes of the lepton flavour-violating processes involving gauge bosons,
like $\mu^- \to e^- \gamma$ and $Z^0 \to e^- \mu^+$,
scale down as $1 \! \left/ m_R^2 \right.$ when $m_R \to \infty$;
this holds even when in those processes
the gauge bosons $\gamma$ and $Z^0$ are virtual,
\textit{i.e.}\ they are off-mass shell;
\item
the amplitudes of the box diagrams for lepton flavour-violating processes
like $\tau^- \to \mu^- \mu^- e^+$ and $\tau^- \to e^- e^- \mu^+$
also scale down as $1 \! \left/ m_R^2 \right.$ for a large seesaw scale;
\item
however, 
if $n \ge 2$,
the amplitudes for lepton flavour-violating processes
$\ell_1^- \to \ell_2^- \left( S^0_b \right)^*$,
where $\left( S^0_b \right)^\ast$ is a virtual (off-mass shell) neutral scalar,
approach a nonzero limit when $m_R \to \infty$.
The non-decoupling of the seesaw scale
in $\ell_1^- \to \ell_2^- \left( S^0_b \right)^*$
is an effect of the one-loop diagrams with neutrinos and charged scalars
in the loop.
\end{enumerate}
As a consequence,
in our framework the amplitude of the process $\mu^- \to e^- e^+ e^-$, 
which derives from $\mu^- \to e^- \left( S^0_b \right)^*$ 
followed by $\left( S^0_b \right)^* \to e^+ e^-$,
is \emph{unsuppressed}\/ in the limit $m_R \to \infty$.
The same happens to the amplitudes of the four $\tau^-$ decays of the same type.

It is important to stress that in our model
the amplitude for $\mu^- \to e^- e^+ e^-$ is unsuppressed
because of the penguin diagrams for neutral-scalar emission
in the $\mu^- \to e^-$ conversion;
indeed,
the penguin diagrams for either $\gamma$ or $Z^0$ emission
vanish in the limit $m_R \to \infty$.
Thus,
our model for lepton-flavour violation differs from,
for instance,
the scotogenic model discussed in ref.~\cite{3840},
wherein it is precisely the $\gamma$ and $Z^0$ penguins
that are instrumental in $\mu^- \to e^- e^+ e^-$
and in muon--electron conversion in nuclei.\footnote{In this paper
  we do not address muon--electron conversion in nuclei
  because in order to do it we would need to specify,
  through \emph{additional assumptions},
  the Yukawa couplings of the extra Higgs doublets to the quarks.
  This is so because in our model muon--electron conversion in nuclei
  occurs---in the limit $m_R \to \infty$---through
  $\mu^- \to e^- \left( S^0_b \right)^\ast$ followed by
  the $\left( S^0_b \right)^\ast$ coupling to quarks.}

Let us estimate a lower bound on $m_R$ by using the experimental bounds,
given in table~\ref{bounds},\footnote{Two new experiments
are planned in search for lepton flavour-violation
at the Paul-Scherrer Institute.
The MEG~II experiment~\cite{baldini}
plans a sensitivity improvement of one order of magnitude 
for $\mu^+ \to e^+ \gamma$.
The $Mu3e$ experiment~\cite{blondel},
which is in the stage of construction, 
aims at a sensitivity for $\mbox{BR} \left( \mu^+ \to e^+ e^- e^+ \right)$
of order $10^{-16}$.}
on the radiative decays $\ell_1 \to \ell_2 \gamma$. 
\begin{table}
\begin{center}
\begin{tabular}{rcl}
$\mbox{BR} \left( \mu^+ \to e^+ \gamma \right)$ &$<$& $4.2 \times 10^{-13}$ \\
$\mbox{BR} \left( \tau^- \to e^- \gamma \right)$ &$<$& $3.3 \times 10^{-8}$  \\
$\mbox{BR} \left( \tau^- \to \mu^- \gamma \right)$ &$<$& $4.4 \times 10^{-8}$ \\
$\mbox{BR} \left( \mu^- \to e^- e^+ e^- \right)$ &$<$& $1.0 \times 10^{-12}$ \\
$\mbox{BR} \left( \tau^- \to e^- e^+ e^- \right)$ &$<$& $2.7 \times 10^{-8}$ \\
$\mbox{BR} \left( \tau^- \to e^- \mu^+ \mu^- \right)$ &$<$&
  $2.7 \times 10^{-8}$ \\
$\mbox{BR} \left( \tau^- \to \mu^- \mu^+ \mu^- \right)$ &$<$&
  $2.1 \times 10^{-8}$ \\
  $\mbox{BR} \left( \tau^- \to \mu^- e^+ e^- \right)$ &$<$& $1.8 \times 10^{-8}$
  \\
\end{tabular}
\end{center}
\caption{The experimental bounds on
  the branching ratios of some lepton flavour-changing decays.
  All the bounds are at the 90\%~CL.
  The first bound is from ref.~\cite{meg},
  all the other bounds are from ref.~\cite{rpp}.
  \label{bounds}}
\end{table}
The amplitude for any such decay has the form
\begin{equation}\label{Agamma}
\mathcal{A} \left( \ell_1^\pm \to \ell_2^\pm \gamma \right)
= e\, \varepsilon_\rho^\ast\,
\bar u_2 \left( i \sigma^{\rho \lambda} q_\lambda \right) 
\left( A_L \gamma_L + A_R \gamma_R \right) u_1,
\end{equation}
where $\varepsilon_\rho$ is the polarization vector of the photon,
$u_1$ and $u_2$ are the spinors of $\ell_1^\pm$ and $\ell_2^\pm$,
respectively,
and $\gamma_L$ and $\gamma_R$ are the projectors of chirality.
The decay rate is given,
in the limit $m_{\ell_2} = 0$,
by
\begin{equation}
  \Gamma \left( \ell_1^\pm \to \ell_2^\pm \gamma \right)
  = \frac{\alpha m_{\ell_1}^3}{4} \left( \left| A_L \right|^2 +  
\left| A_R \right|^2  \right).
\end{equation}
Knowing that $A_L$ and $A_R$ are suppressed by $m_R^{-2}$,
one may estimate,
just on dimensional grounds,
that
\begin{equation}
A_{L,R} \sim \frac{1}{16 \pi^2}\, \frac{m_{\ell_1}}{m_R^2}.
\end{equation}
Using the first two bounds of table~\ref{bounds}
together with the experimental values
for the masses and widths of the $\mu$ and $\tau$,
one may then derive the lower bounds
$m_R \gtrsim 50$\,TeV
from $\mu^+ \to e^+ \gamma$
and $m_R \gtrsim 2$\,TeV from $\tau^- \to e^- \gamma$. 

Thus,
in the framework of ref.~\cite{GL2002},
if we take $m_R \gtrsim 500$\,TeV
then the radiative decays $\ell_1 \to \ell_2 \gamma$ 
are invisible in the foreseeable future. 
On the other hand,
because of the nonzero limit of the amplitudes
for $\ell_1 \to \ell_2 \left( S^0_b \right)^*$,
the charged-lepton decays $\ell_1 \to \ell_2 \ell_3^+ \ell_3^-$
are unsuppressed when $m_R \to \infty$.
It is the purpose of this paper to investigate those decays numerically
in the framework of ref.~\cite{GL2002},
\emph{assuming $m_R$ to be so large that
the radiative charged-lepton decays are invisible}.
Then,
$m_R$ is also much larger than the masses of the scalars in the model,
which we assume to be in between one and a few TeV.

As a sideline,
in this paper we also consider
the contributions of both the neutral and charged scalars
to the anomalous magnetic moment $a_\ell$ of the charged lepton $\ell$,
with particular emphasis on $a_\mu$.

In order to keep the number of parameters of the model at a minimum,
we restrict ourselves to just two Higgs doublets. 
Anticipating our results,
we find that \emph{all}\/ five decays $\ell_1 \to \ell_2 \ell_3^+ \ell_3^-$ 
may well be just around the corner,
while at the same time the contributions
of the non-Standard Model (SM) scalars of the model
can make up for the discrepancy $a_\mu^\mathrm{exp} - a_\mu^\mathrm{SM}$
of the anomalous magnetic moment of the muon.

This paper is organized as follows.
In section~\ref{LFV} we recall some results of ref.~\cite{GL2002}.
We then specialize to the case of just two Higgs doublets in section~\ref{2HD}.
We present the formulas for the contribution
of the non-SM scalars to $a_\ell$ in section~\ref{MM}.
Section~\ref{numerics} is devoted to a numerical simulation.
In section~\ref{concl} we summarize and conclude.

\section{The lepton
  flavour-violating decays $\ell_1^- \to \ell_2^- \ell_3^+ \ell_3^-$} 
\label{LFV}

\subsection{The effective lepton flavour-violating interaction}
\label{2.1}

The framework of ref.~\cite{GL2002} assumes an $n$-Higgs-doublet setup
wherein the violation of the family lepton numbers $L_\ell$ is soft.
The corresponding Yukawa Lagrangian has the form
\begin{equation}
{\cal L}_{\rm Yukawa} = - \sum_{k=1}^{n}\, \sum_{\ell = e, \mu, \tau}
\left[ \phi_k^\dagger\, \bar \ell_R \left( \Gamma_k \right)_{\ell \ell}
+ {\tilde \phi_k}^\dagger\, \bar \nu_{\ell R}
\left( \Delta_k \right)_{\ell \ell} \right] D_{\ell L} + {\rm H.c.}
\label{Yukawa}
\end{equation}
The basic assumption is
\begin{equation}
\mbox{the\ matrices}\ \Gamma_k \; \mbox{and}\ \Delta_k\; \mbox{are\ diagonal},\
\forall \; k=1,\ldots,n,
\end{equation}
as is already implicit in equation~\eqref{Yukawa}.
In that equation,
the Higgs doublets and the left-handed-lepton gauge doublets are given by 
\begin{equation}
  \phi_k = \left( \begin{array}{c} \varphi_k^+ \\*[0.5mm] \varphi_k^0
  \end{array} \right),
  \quad
  \tilde \phi_k = \left( \begin{array}{c} {\varphi_k^0}^\ast
    \\*[0.5mm] - \varphi_k^- \end{array} \right),
  \quad \mbox{and} \quad
  D_{\ell L} = \left( \begin{array}{c} \nu_{\ell L} \\*[0.5mm] \ell_L 
  \end{array} \right),
\end{equation}
respectively.

The scalar mass eigenfields $S_a^+$ and $S_b^0$
are related to the $\varphi_k^+$ and $\varphi^0_k$ by
\begin{equation}
  \varphi_k^+ = \sum_{a=1}^n U_{ka} S^+_a
  \quad \mbox{and} \quad
  \varphi_k^0 =  \frac{1}{\sqrt{2}} \left( v_k + \sum_{b=1}^{2n} V_{kb}
  S_b^0 \right),
  \label{guywo}
\end{equation}
respectively~\cite{osland}.
The vacuum expectation values (VEVs) are $v_k \left/ \sqrt{2} \right.$.
The unitary $n \times n$ matrix $U$ diagonalizes the Hermitian mass
matrix of the charged scalars. 
The $2n \times 2n$ real orthogonal matrix $\tilde V$,
which diagonalizes the mass matrix of neutral scalar
fields, is written as~\cite{osland}
\begin{equation}
  \label{bjghw}
  \tilde V = \left( \begin{array}{c} \mathrm{Re}\,V \\ \mathrm{Im}\, V
  \end{array} \right)
  \quad \mbox{with} \quad
  V \equiv \mathrm{Re}\,V + i\, \mathrm{Im}\, V.
\end{equation}
The matrix $V$ is $n \times 2n$.
We number the scalar mass eigenfields
in such a way that $S_1^\pm = G^\pm$ and $S_1^0 = G^0$ are the Goldstone bosons.  
If there is only one Higgs doublet,
\textit{i.e.}\ when $n=1$,
the matrix $V$ is simply $V = \left( i,\ 1 \right)$
in the phase convention where $v_1 > 0$, 
and $S^0_2$ is the Higgs field of the SM.

We define the diagonal matrices
\be
\label{gfytu}
M_D = \sum_{k=1}^n \frac{v_k}{\sqrt{2}}\, \Delta_k,
\quad
M_\ell = \sum_{k=1}^n \frac{v_k^\ast}{\sqrt{2}}\, \Gamma_k
= \mathrm{diag} \left( m_e, m_\mu, m_\tau \right).
\ee

According to ref.~\cite{GL2002},
in the limit $m_R \to \infty$,
where $m_R$ is the seesaw scale,
the flavour-changing interactions of the physical neutral scalars $S^0_b$,
induced by loops with charged scalars and neutrinos,
are given by 
\begin{equation}
\label{Leff}
\mathcal{L}_\mathrm{eff} \left( S^0 \right) = \sum_{b \ge 2} S^0_b\,
\sum_{\ell_1 \neq \ell_2} 
\bar \ell_1 \left[ 
\left( A_L^b \right)_{\ell_1 \ell_2} \gamma_L + 
\left( A_R^b \right)_{\ell_1 \ell_2} \gamma_R \right] \ell_2.
\end{equation}
Note that the summation over $b$ begins with $b=2$,
\textit{i.e.}\ it excludes the Goldstone boson $S_1^0$.
The coefficients $\left( A_{L,R}^b \right)_{\ell_1 \ell_2}$
were computed in ref.~\cite{GL2002}.
Let us define the $3 \times 3$ unitary matrix $U_R$
that
diagonalizes
$M_R$ as
\begin{equation}\label{RRR}
U_R^\dagger M_R U_R^* = \diag \left( m_4, m_5, m_6 \right),
\end{equation}
where $m_{4,5,6}$ are,
in the limit $m_R \to \infty$,
the masses of the heavy neutrinos.
We next define
\be\label{X}
X_{\ell_1 \ell_2} \equiv \frac{1}{16 \sqrt{2} \pi^2}\, \sum_{i=4}^6 \left[
\left( U_R \right)_{\ell_1 i} \left( U_R^\ast \right)_{\ell_2 i}
\ln{\frac{m_i^2}{\mu^2}} \right] = X_{\ell_2 \ell_1}^\ast,
\ee
where $\mu$ is a mass scale which is arbitrary because
of the unitarity of $U_R$.
Finally,
we define the flavour space matrices $A_{1,2,3}$ as
\bs
\label{A123}
\ba
\left( A_1 \right)_{\ell_1 \ell_2} &\equiv&
\sum_{k=1}^n \left( \Gamma_k \right)_{\ell_1 \ell_1}
\left( \Delta_k \right)_{\ell_2 \ell_2},
\\
\left( A_2 \right)_{\ell_1 \ell_2} &\equiv&
\sum_{k=1}^n \left( \Delta_k^\ast \right)_{\ell_1 \ell_1}
\left( \Delta_k \right)_{\ell_2 \ell_2},
\\
\left( A_3 \right)_{\ell_1 \ell_2} &\equiv&
\sum_{k=1}^n \left( \Delta_k^\ast \right)_{\ell_1 \ell_1}
\left( \Gamma_k^\ast \right)_{\ell_2 \ell_2}.
\ea
\es
Notice that $A_3 = A_1^\dagger$ and $A_2 = A_2^\dagger$.
Then,
\be
\left( A_L^b \right)_{\ell_1 \ell_2} = \frac{X_{\ell_1 \ell_2} A^b_{\ell_1 \ell_2}}
     {m_{\ell_1}^2 - m_{\ell_2}^2}
     \quad \mbox{and} \quad
     \left( A_R^b \right)_{\ell_1 \ell_2} =
     \frac{X_{\ell_2 \ell_1}^\ast \left( A^b_{\ell_2 \ell_1} \right)^\ast}
{m_{\ell_2}^2 - m_{\ell_1}^2},
\ee
where $m_{\ell_i}$ is the mass of the charged lepton $\ell_i$ and
\ba
A^b_{\ell_1 \ell_2} &=& \sum_{k=1}^n V_{kb}^\ast \left\{
\left( \Delta_k^\ast \right)_{\ell_1 \ell_1} \left( m_{\ell_1}^2 - m_{\ell_2}^2 \right)
\left( A_1 \right)_{\ell_1 \ell_2}
\right. \no & &
+ \left( \Gamma_k \right)_{\ell_1 \ell_1} \left[
- m_{\ell_1} \left( M_D^\ast \right)_{\ell_1 \ell_1} \left( A_1 \right)_{\ell_1 \ell_2}
+ \frac{m_{\ell_2}^2}{2}\, \left( A_2 \right)_{\ell_1 \ell_2}
- m_{\ell_2} \left( M_D \right)_{\ell_2 \ell_2} \left( A_3 \right)_{\ell_1 \ell_2}
\right]
\no & & \left.
+ \left( \Gamma_k \right)_{\ell_2 \ell_2} \left[
m_{\ell_2} \left( M_D^\ast \right)_{\ell_1 \ell_1} \left( A_1 \right)_{\ell_1 \ell_2}
  - \frac{m_{\ell_1} m_{\ell_2}}{2}\, \left( A_2 \right)_{\ell_1 \ell_2}
+ m_{\ell_1} \left( M_D \right)_{\ell_2 \ell_2} \left( A_3 \right)_{\ell_1 \ell_2}
\right] \right\}.
\no & &
\label{AAAA}
\ea

We note that,
in every multi-Higgs-doublet model,
it is possible to choose a basis for the scalar doublets
such that only one of them,
say $\phi_1$,
has nonzero VEV:
\be
\left\langle \varphi_1^0 \right\rangle_0 = \frac{v}{\sqrt{2}},
\quad
\left\langle \varphi_k^0 \right\rangle_0 = 0 \quad \forall \; k > 1.
\ee
This basis is called the `Higgs basis'.
In it,
from equation~\eqref{gfytu},
\be
\label{dsifp}
\left( \Delta_1^\ast \right)_ {\ell_1 \ell_1}
= \frac{\sqrt{2}}{v^\ast} \left( M_D^\ast \right)_{\ell_1 \ell_1},
\quad
\left( \Gamma_1 \right)_{\ell_1 \ell_1}
= \frac{\sqrt{2}}{v^\ast}\, m_{\ell_1}.
\ee
With equations~\eqref{dsifp} one finds that,
in the sum over $k$ in equation~\eqref{AAAA},
the term with $k=1$ gives a null contribution.
Thus,
\emph{in the Higgs basis,
the contribution to $A^b_{\ell_1 \ell_2}$
proportional to $V_{1b}^\ast$ is identically zero}.
In particular,
if there is only one Higgs doublet,
\textit{i.e.}\ in the SM,
$A^b_{\ell_1 \ell_2} = 0$,
\textit{viz.}\ when $n=1$ there are no
effective lepton flavour-violating interactions
of the neutral scalar in the limit $m_R \to \infty$.

\subsection{The decay rate}

If $\ell_2 \neq \ell_3$, 
then $\ell_1^- \to \ell_2^- \ell_3^+ \ell_3^-$
may be either $\tau^- \to \mu^- e^+ e^-$ or $\tau^- \to e^- \mu^+ \mu^-$.
Equation~(\ref{Leff}) supplies the amplitude of the subprocess
$\ell_1^- \to \ell_2^- \left( S^0_b \right)^*$.
For the subsequent $\left( S^0_b \right)^* \to \ell_3^+ \ell_3^-$
we have
\begin{equation}
  \label{Leff2}
\mathcal{L}^{(\ell^\pm)}_\mathrm{Yukawa} \left( S^0 \right) = -\frac{1}{\sqrt{2}}\,
\sum_{b=2}^{2n} S^0_b
\sum_{\ell=e,\mu\tau} 
\bar \ell \left[ \left( \hat\Gamma_b \right)_{\ell \ell} \gamma_L
+ \left( {\hat\Gamma}_b^\dagger \right)_{\ell \ell} \gamma_R \right] \ell,
\end{equation}
where
\be \label{vugop}
\hat\Gamma_b \equiv \sum_{k=1}^n V_{kb}^* \Gamma_k.
\ee
We write the decay amplitude for $\ell_1^- \to \ell_2^- \ell_3^+ \ell_3^-$ as
\begin{equation}
\label{ampl}
\mathcal{A} = \sum_{b=2}^{2n}
\bar u_2 \left[ \left( \lambda_b \right)_{\ell_2 \ell_1} \gamma_L +
\left( \rho_b \right)_{\ell_2 \ell_1} \gamma_R \right] u_1\,
\bar u_3 \left[ \left( \hat \Gamma_b \right)_{\ell_3 \ell_3} \gamma_L
+ \left( \hat \Gamma_b^\dagger \right)_{\ell_3 \ell_3} \gamma_R \right] v_3,
\end{equation}
where,
from equations~\eqref{Leff} and~\eqref{Leff2},
\begin{equation}
  \label{vytop}
  \left( \lambda_b \right)_{\ell_2 \ell_1} = 
- \frac{\left( A_L^b \right)_{\ell_2 \ell_1}}{\sqrt{2}\, M_b^2}
\quad \mbox{and} \quad
\left( \rho_b \right)_{\ell_2 \ell_1} = 
- \frac{\left( A_R^b \right)_{\ell_2 \ell_1}}{\sqrt{2}\, M_b^2}.
\end{equation}
In equations~\eqref{vytop},
$M_b$ is the mass of $S^0_b$.
In the scalar propagators,
we have neglected the four-momentum of the $\ell_3^+ \ell_3^-$ subsystem.
With the amplitude in equation~(\ref{ampl}),
the decay rate is given by
\begin{eqnarray}
\Gamma \left( \ell_1^- \to \ell_2^- \ell_3^+ \ell_3^- \right) &=& 
\frac{m_{\ell_1}^5}{6144 \pi^3}
\left[ \left| \sum_{b=2}^{2n} \left( \lambda_b \right)_{\ell_2 \ell_1}
\left( {\hat \Gamma}_b \right)_{\ell_3 \ell_3} \right|^2
+ \left| \sum_{b=2}^{2n} \left( \lambda_b \right)_{\ell_2 \ell_1}
\left( {\hat \Gamma}_b^\ast \right)_{\ell_3 \ell_3} \right|^2
\right. \no & & \left.
+ \left| \sum_{b=2}^{2n} \left( \rho_b \right)_{\ell_2 \ell_1}
\left( {\hat \Gamma}_b \right)_{\ell_3 \ell_3} \right|^2 +
\left| \sum_{b=2}^{2n} \left( \rho_b \right)_{\ell_2 \ell_1}
\left( {\hat \Gamma}_b^\ast \right)_{\ell_3 \ell_3} \right|^2 
\right].
\label{rate1}
\end{eqnarray}
We have neglected the masses of the final charged leptons in the kinematics.

If $\ell_2 = \ell_3$, 
then $\ell_1^- \to \ell_2^- \ell_2^+ \ell_2^-$
may be either $\mu^- \to e^- e^+ e^-$ 
or $\tau^- \to e^- e^+ e^-$ or $\tau^- \to \mu^- \mu^+ \mu^-$.
In equation~(\ref{ampl}) one must antisymmetrize the amplitude with respect
to $\ell_2^-$ and in the kinematics one must insert an extra factor $1/2$.
The final result is
\begin{eqnarray}
\Gamma \left( \ell_1^- \to \ell_2^- \ell_2^+ \ell_2^- \right) &=& 
\frac{m_{\ell_1}^5}{6144 \pi^3} \left[\,
\frac{1}{2}\,
\left| \sum_{b=2}^{2n} \left( \lambda_b \right)_{\ell_2 \ell_1}
\left( {\hat \Gamma}_b \right)_{\ell_2 \ell_2} \right|^2
+ \frac{1}{2}\,
\left| \sum_{b=2}^{2n} \left( \rho_b \right)_{\ell_2 \ell_1}
\left( {\hat \Gamma}_b^\ast \right)_{\ell_2 \ell_2} \right|^2 
\right. \no & & \left.
+ 
\left| \sum_{b=2}^{2n} \left( \lambda_b \right)_{\ell_2 \ell_1}
\left( {\hat \Gamma}_b^\ast \right)_{\ell_2 \ell_2} \right|^2
+ 
\left| \sum_{b=2}^{2n} \left( \rho_b \right)_{\ell_2 \ell_1}
\left( {\hat \Gamma}_b \right)_{\ell_2 \ell_2} \right|^2
\right].
\end{eqnarray}

\section{Two Higgs doublets}
\label{2HD}

From now on we assume $n=2$,
\textit{i.e.}\ a two-Higgs-doublet model.
In the Higgs basis,
the VEVs are given by
\begin{equation}
  \left\langle \varphi_1^0 \right\rangle_0 = \frac{v}{\sqrt{2}}, \quad
  \left\langle \varphi_2^0 \right\rangle_0 = 0,
\end{equation}
where $v \approx 246\,$GeV is real and positive.
Thus,
according to equation~\eqref{guywo},
\begin{equation}
  \label{fptmjk}
  \varphi_k^0 =  \frac{1}{\sqrt{2}} \left( \delta_{k1} v + \sum_{b=1}^4 V_{kb}
  S_b^0 \right). 
\end{equation}
Moreover,
the matrix $U$ is the $2 \times 2$ unit matrix,
\textit{i.e.}\ $\varphi_1^+ = S_1^+ = G^+$ is the charged Goldstone boson
and $\varphi_2^+ = S_2^+$ is the physical charged scalar.
According to the notation of ref.~\cite{osland},
the $4 \times 4$ orthogonal matrix $\tilde V$ of equation~\eqref{bjghw},
which diagonalizes the mass matrix of neutral scalar fields,
is given by
\begin{equation}
  \tilde V =
  \left( \begin{array}{cccc}
    0 & R_{11} & R_{12} & R_{13} \\
    0 & R_{21} & R_{22} & R_{23} \\
    1 & 0 & 0 & 0 \\
    0 & R_{31} & R_{32} & R_{33} 
  \end{array} \right),
  \quad \textit{i.e.}\
  V = \left( \begin{array}{cccc}
    i & R_{11} & R_{12} & R_{13} \\
    0 & R_{21} + i R_{31} & R_{22} + i R_{32} & R_{23} + i R_{33}
  \end{array} \right),
\end{equation}
with a $3 \times 3$ orthogonal matrix $R$.
The third row of $\tilde V$
corresponds to the neutral Goldstone boson $S_1^0 = G^0$.
The definition~\eqref{vugop} reads
\be
\label{fogpy}
\hat \Gamma_b = \Gamma_1 R_{1\,b-1} + 
\Gamma_2 \left( R_{2\,b-1} - i R_{3\,b-1} \right)
\ee
for $b = 2, 3, 4$.
We parameterize the flavour-diagonal Yukawa coupling matrices as 
\begin{subequations}\label{2hdys}
  \begin{eqnarray}
    \Gamma_1 &=& \frac{\sqrt{2}}{v}\,
    \diag \left( m_e, m_\mu, m_\tau \right) =
    \frac{\sqrt{2}}{v}\, M_\ell, \\
    \Gamma_2 &=& 
    \diag \left( \gamma_e, \gamma_\mu, \gamma_\tau \right), \\
    \Delta_1 &=& \diag \left( d_e, d_\mu, d_\tau \right) =
    \frac{\sqrt{2}}{v}\,M_D, \\
    \Delta_2 &=& \diag \left( \delta_e, \delta_\mu, \delta_\tau \right).
  \end{eqnarray}
\end{subequations}
Therefore,
from equations~\eqref{A123} and~\eqref{AAAA},
\bs
\ba
A^b_{\ell_1 \ell_2} &=& V_{2b}^\ast\, A_{\ell_1 \ell_2},
\label{cuigp} \\
A_{\ell_1 \ell_2} &=&
\frac{\sqrt{2} \left( m_{\ell_1}^2 - m_{\ell_2}^2 \right)
  m_{\ell_1} \delta_{\ell_1}^* d_{\ell_2}}
{v}
\no & &
- \left( m_{\ell_1}^2 + \frac{m_{\ell_2}^2}{2} \right)
\gamma_{\ell_1} d_{\ell_1}^* d_{\ell_2}
+ \frac{3 m_{\ell_1} m_{\ell_2}}{2}\, d_{\ell_1}^* \gamma_{\ell_2} d_{\ell_2}
\no & &
+ \left( m_{\ell_1}^2 - \frac{m_{\ell_2}^2}{2} \right)
\gamma_{\ell_1} \delta_{\ell_1}^* \delta_{\ell_2}
- \frac{m_{\ell_1} m_{\ell_2}}{2}\, \delta_{\ell_1}^* \gamma_{\ell_2} \delta_{\ell_2}
\no & &
+ \frac{v m_{\ell_2}}{\sqrt{2}}
\left( \gamma_{\ell_1} d_{\ell_1}^* \gamma_{\ell_2} \delta_{\ell_2}
- \gamma_{\ell_1} \delta_{\ell_1}^* \gamma_{\ell_2}^* d_{\ell_2} \right)
+ \frac{v m_{\ell_1}}{\sqrt{2}}
\left( \delta_{\ell_1}^* \left| \gamma_{\ell_2} \right|^2 d_{\ell_2}
- \gamma_{\ell_1}^2 d_{\ell_1}^* \delta_{\ell_2} \right).
\hspace*{5mm}
\label{hlrpsa}
\ea
\es
As demonstrated at the end of section~\ref{2.1},
in $A^b_{\ell_1 \ell_2}$ the term proportional to $V_{1b}^\ast$ vanishes.

We now make the further assumption that
$\phi_1$ is just identical with the Higgs doublet of the SM;
this means that $S^0_2$ is exactly like the SM Higgs boson.
This choice relieves us from having to take into account
the experimental restrictions on the couplings of the SM Higgs boson,
which become automatically fulfilled.
We now have
\be
\phi_1 = \left( \begin{array}{c} S_1^+ \\
\left( v + S_2^0 + i S_1^0 \right) \left/ \sqrt{2} \right.
\end{array} \right),
\ee
where $S_1^+ = G^+$ and $S_1^0 = G^0$ are the Goldstone bosons.
This means that we choose $R_{11} = 1$,
whence it follows that $R$ can be written as\footnote{We assume
  without loss of generality that the orthogonal matrix $R$
  has determinant $+1$.}
\begin{equation}\label{R}
  R = \left(
  \begin{array}{ccc}
    1 & 0 & 0 \\
    0 & \cos{\alpha} & \sin{\alpha} \\
    0 & -\sin{\alpha} & \cos{\alpha}
  \end{array} \right).
\end{equation}
The matrix $V$ is
\be
V = \left( \begin{array}{cccc} i & 1 & 0 & 0 \\
  0 & 0 & e^{- i \alpha} & i e^{- i \alpha} \end{array} \right).
\ee
Thus,
from equation~\eqref{fptmjk},
\be
\label{vigop}
\phi_2 = \left( \begin{array}{c} S_2^+ \\ e^{- i \alpha}
  \left( S_3^0 + i S_4^0 \right) \left/ \sqrt{2} \right.
\end{array} \right).
\ee

From equation~\eqref{fogpy},
\be
\hat \Gamma_2 = \Gamma_1, \quad \hat \Gamma_3 = e^{i \alpha} \Gamma_2,
\quad \hat \Gamma_4 = - i e^{i \alpha} \Gamma_2,
\ee
and,
from equation~\eqref{cuigp},
\be
A^2_{\ell_1 \ell_2} = 0, \quad
A^3_{\ell_1 \ell_2} = e^{i \alpha} A_{\ell_1 \ell_2}, \quad
A^4_{\ell_1 \ell_2} = - i e^{i \alpha} A_{\ell_1 \ell_2}.
\ee
The decay rates are then
\bs
\label{bihpi}
\ba
\Gamma \left( \ell_1^- \to \ell_2^- \ell_3^+ \ell_3^- \right) &=&
\frac{m_{\ell_1}^5}{6144 \pi^3}
\left| X_{\ell_2\ell_1} \right|^2 \left| \gamma_{\ell_3} \right|^2
\frac{\left| A_{\ell_2 \ell_1} \right|^2 + \left| A_{\ell_1 \ell_2} \right|^2}
{\left( m_{\ell_2}^2 - m_{\ell_1}^2 \right)^2}
\left( \frac{1}{M_3^4} + \frac{1}{M_4^4} \right),
\label{r1} \\
\Gamma \left( \ell_1^- \to \ell_2^- \ell_2^+ \ell_2^- \right) &=&
\frac{m_{\ell_1}^5}{6144 \pi^3}
\left| X_{\ell_2\ell_1} \right|^2 \left| \gamma_{\ell_2} \right|^2
\frac{\left| A_{\ell_2 \ell_1} \right|^2 + \left| A_{\ell_1 \ell_2} \right|^2}
{\left( m_{\ell_2}^2 - m_{\ell_1}^2 \right)^2}
\nonumber \\ && \hspace{39mm} \times 
\left[ \frac{3}{4} \left( \frac{1}{M_3^4} + \frac{1}{M_4^4} \right)
+ \frac{1}{2 M_3^2 M_4^2} \right]. \hspace*{8mm}
\label{r2}
\ea
\es
The decay rates depend on the masses $M_3$ and $M_4$
of the non-SM neutral scalar fields $S^0_3$ and $S^0_4$,
respectively. 
There is no dependence on the phase $\alpha$.
In equation~\eqref{r1},
$\ell_2 \neq \ell_3$ is understood.

\section{The anomalous magnetic moment of the muon}
\label{MM}

Let $a_\ell^{(S)}$ denote the contributions of the non-SM scalars $S^0_3$,
$S^0_4$,
and $S^\pm_2$ to the anomalous magnetic moment (AMM)
of the charged lepton $\ell$. 
To a good approximation,
\begin{subequations}
  \label{aell}
\begin{eqnarray}
a_\ell^{(S)} &\simeq&
\frac{m_\ell^2}{96 \pi^2} \left\{
2 \left| \gamma_\ell \right|^2
\left( \frac{1}{M_3^2} + \frac{1}{M_4^2} \right) \right.
\label{a} \\ & &
- 3\, \mbox{Re} \left( e^{2i\alpha} \gamma_\ell^2 \right)
\left[ \frac{1}{M_3^2} \left( 3 + 2 \ln \frac{m_\ell^2}{M_3^2} \right) -
  \frac{1}{M_4^2} \left( 3 + 2 \ln \frac{m_\ell^2}{M_4^2} \right) \right]
\label{b} \\ & & \left. 
- \frac{\left| \gamma_\ell \right|^2}{\mu_2^2} \right\}.
\label{c}
  \end{eqnarray}
\end{subequations}
Lines~(\ref{a}) and~(\ref{b}) derive from
a loop with $\ell$ and either $S_3^0$ or $S_4^0$;
the photon line attaches to $\ell$.
Line~(\ref{c}) comes from a loop with $S_2^\pm$ and light neutrinos,
wherein the external photon attaches to $S_2^\pm$;
in that line,
$\mu_2$ denotes the mass of $S^\pm_2$.
We have dropped all the terms proportional to $m_R^{-2}$,
including in particular the contributions
from the loop with $S_2^\pm$ and heavy neutrinos.
For the coupling of the charged scalars to the charged leptons
we refer the reader to ref.~\cite{GL2002}.

There is a long-standing discrepancy between the experimental value
of the AMM of the muon,
$a_\mu^\mathrm{exp}$,
and the SM theoretical value of that AMM,
$a_\mu^\mathrm{SM}$~\cite{blum}:\footnote{See also ref.~\cite{lindner}
for a recent review.}
\be\label{discrepancy}
  a_\mu^\mathrm{exp} - a_\mu^\mathrm{SM} = \left\{
  \begin{array}{l}
    (287 \pm 80) \times 10^{-11}\ (\mbox{at}\
    3.6\,\sigma)~\cite{davier}, \\
    (261 \pm 78) \times 10^{-11}\ (\mbox{at}\
    3.3\,\sigma)~\cite{hagiwara}.
  \end{array} \right.
\ee
If this discrepancy signals new physics,
then the contributions of the scalars in our model to the AMM of the muon
may be relevant.
Taking for instance $\gamma_\mu^2$ real and $e^{2i\alpha} = 1$,
one has
\be
a_\mu^{(S)} \simeq
- \frac{\gamma_\mu^2}{96 \pi^2} \left[
\frac{m_\mu^2}{M_3^2} \left( 7 + 6 \ln \frac{m_\mu^2}{M_3^2} \right)
- \frac{m_\mu^2}{M_4^2} \left( 11 + 6 \ln \frac{m_\mu^2}{M_4^2} \right)
+ \frac{m_\mu^2}{\mu_2^2} \right].
\label{vugig}
\ee
The right-hand side of equation~\eqref{vugig} is dominated by the
two terms with logarithms.
One readily sees that the terms with $M_4$ and $\mu_2$
give negative contributions to $a_\mu^{(S)}$
(assuming $\gamma_\mu^2$ to be positive),
while the term with $M_3$ gives a positive contribution;
since $a_\mu^\mathrm{exp} - a_\mu^\mathrm{SM}$ is positive,
we would like the term with $M_3$ to dominate over the other two;
this is achieved with $M_3 < M_4$.
Taking for instance $M_3 = 1\,$TeV,
$M_4 = \mu_2 = 2$\,TeV,\footnote{Our choice $M_4 = \mu_2$ has the advantage
  that it automatically leads to a zero oblique parameter $T$.
  Indeed,
  in our two-Higgs-doublet model with $R_{11} = 1$,
  \be\nonumber
  T = \frac{1}{16 \pi s_w^2 m_W^2} \left[ f \left( M_3^2, \mu_2^2 \right)
    + f \left( M_4^2, \mu_2^2 \right)
    - f \left( M_3^2, M_4^2 \right) \right],
  \ee
  where $f \left( x, y \right)$ is a function~\cite{osland,review}
  that is zero when $x = y$.
  Thus, $T = 0$ when $M_4 = \mu_2$.}
and $\gamma_\mu = 1.7$,
we find $a_\mu^{(S)} = 258 \times 10^{-11}$,
which is of the right sign and absolute value
to explain the discrepancy~\eqref{discrepancy}.
We conclude that our model can,
using reasonable parameters,
fill the gap between $a_\mu^\mathrm{exp}$
and $a_\mu^\mathrm{SM}$.

The experimental AMM of the electron
is in good agreement with the SM prediction for $a_e$.
We must therefore check that the non-SM scalars of our model
give an $a_e^{(S)}$ smaller than the experimental error
$2.6 \times 10^{-13}$~\cite{rpp} of $a_e$.
We might of course simply take $\gamma_e = 0$,
but this would eliminate \textit{e.g.}\ the decay $\mu^- \to e^- e^+ e^-$,
which we would like to have close to its experimental upper limit.
So we use instead the same scalar masses as before
and choose $\gamma_e = 1.7$,
obtaining $a_e^{(S)} = 1.0 \times 10^{-13}$.
Thus,
even for a relatively large $\gamma_e$,
$a_e^{(S)}$ can be below the experimental error.
This is of course because of the tiny electron mass.

\section{Numerics}
\label{numerics}

In this section,
we want to show that in the two-Higgs-doublet version
of the framework of ref.~\cite{GL2002},
and assuming moreover $R_{11} = 1$,
\emph{there is a region in parameter space
  where the branching ratios of all five decays
  $\ell_1^- \to \ell_2^- \ell_3^+ \ell_3^-$
  are close to their present experimental upper bounds}\/
displayed in table~\ref{bounds}.

Notice that we only strive in this section to prove that
something is \emph{possible};
we do \emph{not}\/ attempt a full scan of the parameter space of our model,
which is quite vast.
On the contrary,
we shall make many simplifying assumptions,
for instance \emph{we assume that all the parameters of the model are real}.

In the decay rates of equations~\eqref{bihpi} there are various unknowns:
\begin{enumerate}
\item the neutral-scalar masses $M_3$ and $M_4$;
  \label{p3}
\item the factors $\left| X_{\ell_2 \ell_1} \right|^2$;
  \label{p1}
\item the Yukawa couplings $\gamma_\ell$
  together with those in $A_{\ell \ell'}$.
  \label{p2}
\end{enumerate}
In this section we also want to fit $a_\mu^\mathrm{exp} - a_\mu^\mathrm{SM}$
of equation~(\ref{discrepancy})
by using $a_\mu^{(S)}$ of equation~\eqref{aell};
in that equation there are the neutral-scalar masses $M_3$ and $M_4$,
the charged-scalar mass $\mu_2$,
the Yukawa coupling $\gamma_\mu$,
and the phase $\alpha$.
In order to simplify our task,
\emph{we fix all those parameters at the values used in section~\ref{MM}},
\textit{viz.}
\bs
\label{input1}
\ba
& & M_3 = 1\, \mathrm{TeV}, \quad M_4 = 2\, \mathrm{TeV}, \label{mass}
\\
& & \gamma_\mu = 1.7. \label{gamma}
\ea
\es
Thus,
the neutral-scalar masses mentioned in point~\ref{p3} above
are fixed through equation~\eqref{mass}.
Notice in equation~\eqref{gamma} that $\gamma_\mu$ is assumed to be real.

\begin{figure}[t]
\begin{center}
\epsfig{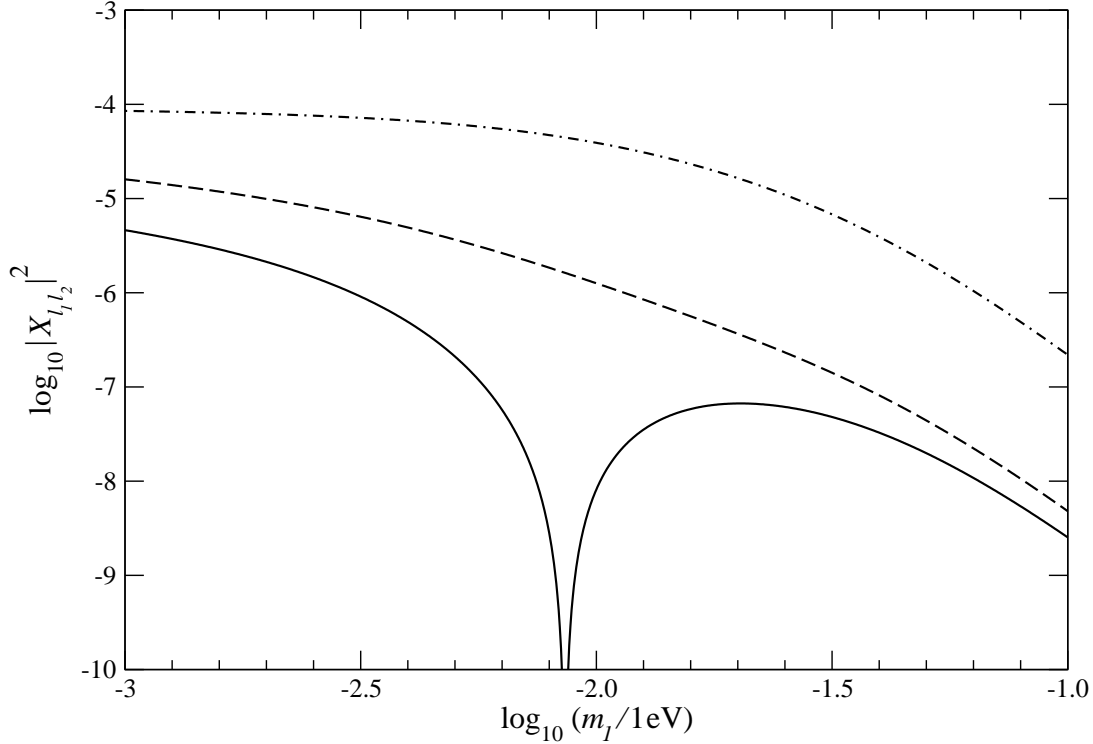}
\end{center}
\caption{The factors $\left| X_{\ell_2 \ell_1} \right|^2$ as functions of $m_1$.
  The full line gives $\left| X_{e \mu} \right|^2$,
  the dashed line gives $\left| X_{e \tau} \right|^2$,
  and the dashed-dotted line is $\left| X_{\mu \tau} \right|^2$.
  \label{XXXX}}
\end{figure}

In order to compute the factors $\left| X_{\ell_2 \ell_1} \right|^2$
we proceed in the following way. 
The mass matrix of the light neutrinos is obtained by the seesaw formula.
In our notation,
it reads
\begin{equation}\label{mnu}
  \mnu = - M_D^T M_R^{-1} M_D
  = -\frac{v^2}{2}\, \Delta_1 M_R^{-1} \Delta_1,
\end{equation}
where $\Delta_1 = \mathrm{diag} \left( d_e, d_\mu, d_\tau \right)$ is diagonal.
We shall fix
\be
\label{input2}
d_e = 0.6, \quad d_\mu = d_\tau = 0.1.
\ee
Inverting equation~(\ref{mnu}),
we obtain
\be
M_R = -\frac{v^2}{2}\, \Delta_1 \mnu^{-1} \Delta_1.
\label{mrr}
\ee
The matrix $\mnu$ is 
diagonalized as
\begin{equation}
V_L^T \mnu V_L = \diag \left( m_1, m_2, m_3 \right) \equiv \hat m,
\label{vl}
\end{equation}
where $m_{1,2,3}$ are the light-neutrino masses
and $V_L = e^{i \hat\alpha}\, U_\mathrm{PMNS}\, e^{i\hat\beta}$
is identical to the lepton mixing matrix $U_\mathrm{PMNS}$,
apart from a diagonal matrix of unphysical phases $e^{i \hat\alpha}$ on the left
and apart from the Majorana phase factors of the diagonal matrix $e^{i\hat\beta}$
on the right.
Using equations~\eqref{mrr} and~\eqref{vl}
together with the fact that the matrices $\Delta_1$,
$\hat m$,
$e^{i \hat \alpha}$,
and $e^{i \hat \beta}$ are diagonal,
we obtain
\be
M_R = -\frac{v^2}{2}\, e^{i \hat\alpha} \Delta_1\, U_\mathrm{PMNS}
\left( e^{2 i \hat\beta} \hat m^{-1} \right)
U_\mathrm{PMNS}^T\, \Delta_1 e^{i\hat\alpha}.
\label{mr}
\ee
Using our simplifying assumption that
all the parameters in the model are real,
we set in equation~\eqref{mr}
$e^{i \hat \alpha} = e^{i \hat \beta} = \mathbbm{1}$
and we also assume that $U_\mathrm{PMNS}$ is real.
Using the standard parameterization for $U_\mathrm{PMNS}$ in ref.~\cite{rpp},
we fix $e^{i\delta} = -1$;\footnote{We might alternatively 
have chosen $e^{i \delta} = +1$;
we have checked that there is no qualitative difference between the two cases.}
we also fix the mixing angles
at their best-fit values of ref.~\cite{schwetz},
\textit{viz.}\ $s_{12}^2 = 0.304$,
$s_{23}^2 = 0.452$,
and $s_{13}^2 = 0.0218$.
We also
have to choose the type of light-neutrino mass spectrum,
either normal or inverted---for definiteness,
we settle on a normal mass spectrum.
Let the lightest neutrino mass $m_1$,
which is unknown to date,
be a free parameter;
with a choice for $m_1$ and the best-fit values
$\Delta m^2_{21} = 7.50 \times 10^{-5}\, \mathrm{eV}^2$
and
$\Delta m^2_{31} = 2.457 \times 10^{-3}\, \mathrm{eV}^2$
of ref.~\cite{schwetz},
we obtain for the other two light-neutrino masses
\be
m_2 = \sqrt{m_1^2 + \Delta m^2_{21}} 
\quad \mbox{and} \quad 
m_3 = \sqrt{m_1^2 + \Delta m^2_{31}}.
\ee
We are now able to compute the matrix $M_R$ as a function of $m_1$
through equation~\eqref{mr};
therefrom we compute the quantities $\left| X_{\ell_2 \ell_1} \right|^2$
by using equations~\eqref{RRR} and~\eqref{X}.
We obtain the result depicted in figure~\ref{XXXX}.
Notice that $X_{e \mu}$ has a zero for $m_1 \approx 0.0086$\,eV;
else,
the $\left| X_{\ell_2 \ell_1} \right|^2$ are decreasing functions of $m_1$,
and vary by a few orders of magnitude
from $m_1 = 0$ to $m_1 = 0.1\,$eV.
From now one we fix
\be
\label{input3}
m_1 = 0.05\,\mbox{eV}.
\ee
We then have
\be
\label{Xfactors}
\left| X_{e \mu} \right|^2 = 1.99 \times 10^{-8}, \quad
\left| X_{e \tau} \right|^2 = 4.43 \times 10^{-8}, \quad
\left| X_{\mu \tau} \right|^2 = 2.11 \times 10^{-6}.
\ee
In this way we have fixed the factors mentioned in point~\ref{p1} above.
Besides equations~\eqref{Xfactors},
we also obtain,
from equation~\eqref{input3},
heavy-neutrino masses $m_4 = 4.3 \times 10^{12}$\,GeV,
$m_5 = 6.0 \times 10^{12}$\,GeV,
and $m_6 = 2.2 \times 10^{14}$\,GeV.
These masses represent the seesaw scale,\footnote{Actually,
  $m_6$ is two orders of magnitude larger than $m_4$ and $m_5$
  and therefore there is no well-defined seesaw scale,
  but that is not relevant for our purposes.}
which is so large that all the radiative charged-lepton decays
are completely invisible.
Actually,
$m_R$ is this large partly because we chose
the Yukawa couplings $d_\ell$ close to one,
\textit{cf.}\ equation~\eqref{input2},
in order to achieve
large $\tau$-lepton branching ratios.\footnote{Note that if one uses
  $d_e = d_\mu = d_\tau = 0$
  in the $A_{\ell_1 \ell_2}$ of equation~\eqref{hlrpsa},
  then only two subdominant terms,
  \textit{i.e.}\ terms without $v$ in the numerator,
  survive.}
Thus,
the effect that we want to produce in our model
can only occur for a large seesaw scale---it disappears,
at least in the case of the $\tau$-lepton,
for small $m_R$.

Some of the Yukawa couplings mentioned in point~\ref{p2}
are given in equations~\eqref{input1} and~\eqref{input2}.
We now fix the remaining Yukawa couplings as
\be
\label{input4}
\gamma_e = \gamma_\tau = 1.7, \quad
\delta_e = 0, \quad
\delta_\mu = 0.00007, \quad
\delta_\tau = 0.2.
\ee
With all these input values,
we obtain the branching ratios
\bs
\label{theBRs}
\ba
\mathrm{BR} \left( \mu^- \to e^- e^+ e^- \right) &=& 3.872 \times 10^{-13},
\\
\mathrm{BR} \left( \tau^- \to e^- e^+ e^- \right) &=& 1.111 \times 10^{-8},
\\
\mathrm{BR} \left( \tau^- \to e^- \mu^+ \mu^- \right) &=& 1.280 \times 10^{-8},
\\
\mathrm{BR} \left( \tau^- \to \mu^- \mu^+ \mu^- \right) &=& 1.307 \times 10^{-8},
\\
\mathrm{BR} \left( \tau^- \to \mu^- e^+ e^- \right) &=& 1.506 \times 10^{-8}.
\ea
\es
%
One sees that all these branching ratios
are less than a factor of three away from the upper bounds
of table~\ref{bounds}.
\emph{We have thus demonstrated that in our model it is possible to
suppress the radiative decays of the muon and tau lepton,
while keeping the branching ratios of their decays into charged leptons
very close to the experimental upper bounds.}

Some remarks concerning the input values that we have utilized are in order:
\begin{itemize}
\item All the experimental upper bounds
  on the branching ratios of the decays
  of the $\tau$-lepton in table~\ref{bounds} are quite similar.
  Therefore,
  if we want to have both $\tau^- \to \ell^- e^+ e^-$
  and $\tau^- \to \ell^- \mu^+ \mu^-$ close to their experimental upper bounds,
  then $\gamma_e$ and $\gamma_\mu$
  will have to be similar---see the explicit factors $\gamma_{\ell_3}$
  and $\gamma_{\ell_2}$ in the decay rates of equations~\eqref{r1}
  and~\eqref{r2},
  respectively.
  For definiteness we have chosen all three $\gamma_\ell$ to be the same.
  In figure~\ref{figure1} we depict the way the five branching ratios vary
  as functions of some $\gamma_\ell$.
\begin{figure}[t]
  \begin{center}
    \includegraphics[scale=0.9]{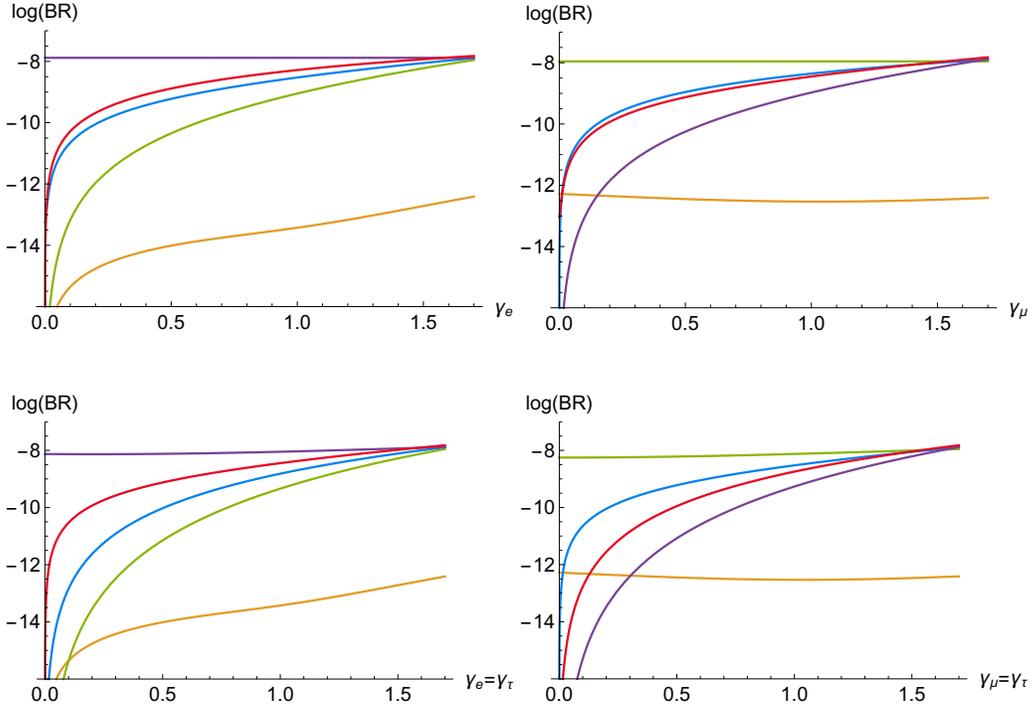}
\end{center}
\caption{$\mathrm{BR} \left( \mu^- \to e^- e^+ e^- \right)$ (orange line),
  $\mathrm{BR} \left( \tau^- \to e^- e^+ e^- \right)$ (green line),
  $\mathrm{BR} \left( \tau^- \to e^- \mu^+ \mu^- \right)$ (blue line),
  $\mathrm{BR} \left( \tau^- \to \mu^- \mu^+ \mu^- \right)$ (lilac line),
  and $\mathrm{BR} \left( \tau^- \to \mu^- e^+ e^- \right)$ (red line)
  as functions of various Yukawa couplings.
  In the top-left figure,
  $\gamma_e$ varies in between 0 and 1.7.
  In the top-right figure,
  $\gamma_\mu$ varies.
  Bottom left,
  $\gamma_e$ and $\gamma_\tau$ change
  but with $\gamma_\tau$ remaining equal to $\gamma_e$.
  In the bottom right,
  $\gamma_\mu = \gamma_\tau$  varies.
  In all the figures,
  all the Yukawa couplings that do not vary take the values
  in equations~\eqref{gamma},
  \eqref{input2},
  and~\eqref{input4}.
  \label{figure1}}
\end{figure}
\item In $A_{\ell_1 \ell_2}$ in equation~\eqref{hlrpsa}
  the dominant terms have $v \simeq 246$\,GeV in the numerator.
  For large $\gamma_e = \gamma_\mu = 1.7$ and large $d_e = 0.6$
  and $d_\mu = 0.1$,
  these terms will give a much too large contribution
  to $\mbox{BR} \left(\mu^- \to e^- e^+ e^-\right)$
  unless there is a delicate cancellation
  between the terms proportional to $\delta_e$
  and the terms proportional to $\delta_\mu$.
  This cancellation is illustrated in figure~\ref{figure2}
  for $\delta_\mu$ of equation~\eqref{input4}.
  For larger values of $\delta_\mu$
  the curve is basically identical but shifted to the right.
\begin{figure}[t]
  \begin{center}
    \includegraphics[scale=0.6]{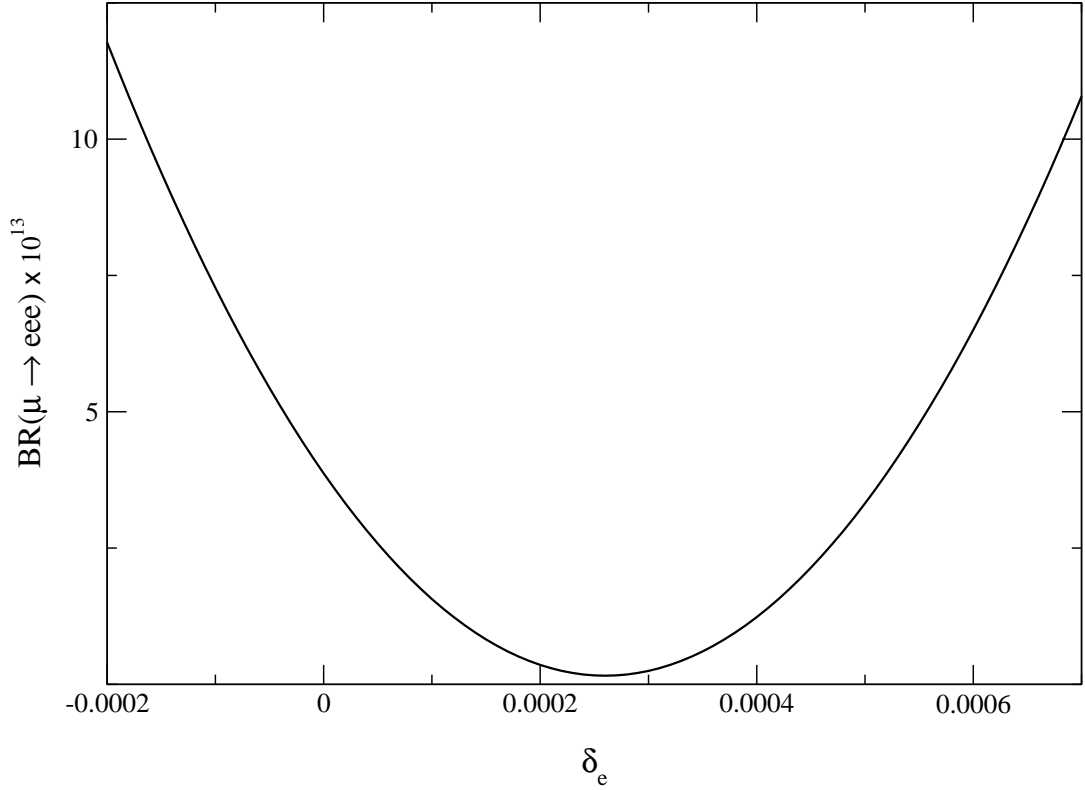}
  \end{center}
  \caption{$\mathrm{BR} \left( \mu^- \to e^- e^+ e^- \right)$
    as a function of the Yukawa coupling $\delta_e$
    while all the other Yukawa couplings remain fixed at their values
    in equations~\eqref{gamma},
    \eqref{input2},
    and~\eqref{input4}.
    The experimental bound
    $\mathrm{BR} \left( \mu^- \to e^- e^+ e^- \right) \times 10^{13} < 10$
    is not depicted in this figure but must be taken into account.
    \label{figure2}}
\end{figure}
\item On the other hand,
in the decays of the $\tau$-lepton the terms with $v$ in the numerator
are just the relevant ones and we have needed,
since we have chosen tiny $\delta_e$ and $\delta_\mu$,
large parameters $\delta_\tau$,
$d_e$,
$d_\mu$,
and $\gamma_\ell$ ($\ell=e,\mu,\tau$).
\end{itemize}
We may thus say that the branching ratios in equations~\eqref{theBRs}
involve some finetuning.

\section{Conclusions} \label{concl}

It is now known,
since the experimental observation of neutrino
oscillations~\cite{nuosc-exp},\footnote{For reviews
  on the phenomenology of neutrino oscillations,
  see for instance ref.~\cite{nuosc-reviews}.}
that there is lepton flavour-violation.
However,
that violation has not yet been observed in the charged-lepton sector
and it is not quite certain where it is most likely to be observed first.
In this context,
the radiative decays $\ell_1^\pm \to \ell_2^\pm \gamma$ seem the best guess,
and decays of the form $\ell_1^\pm \to \ell_2^\pm \ell_3^+ \ell_3^-$
may be an option as well.

In this paper we have demonstrated,
through an explicit numerical example,
that there is a class of models where the 
radiative
decays in the paragraph above
may be so suppressed as to be utterly invisible,
yet any of the five decays
of the form $\ell_1^\pm \to \ell_2^\pm \ell_3^+ \ell_3^-$,
or indeed---if one assumes some finetuning---all such five decays
simultaneously,
may be just around the corner.

Our class of models,
first considered in ref.~\cite{soft},
has three right-handed neutrino singlets and has more than one Higgs doublet.
The crucial assumption is that the lepton flavours are conserved
in the Yukawa couplings
and broken only in the Majorana mass terms of the right-handed neutrinos;
this assumption is field-theoretically consistent
because those mass terms have dimension three
while the Yukawa couplings have dimension four.
As demonstrated in ref.~\cite{GL2002},
the effect mentioned in the previous paragraph occurs if the seesaw scale is
much larger than all other scales in this class of models. 
In the present paper we have shown that 
there is a relevant simplification
of the effective flavour-violating couplings of the neutral scalars, 
emerging at the one-loop level,
when one uses the Higgs basis,
\textit{i.e.}\ the basis for the Higgs doublets wherein
only one of them has nonzero VEV.

We have explicitly computed the branching ratios
of the five decays $\ell_1^\pm \to \ell_2^\pm \ell_3^+ \ell_3^-$
in the case of a two-Higgs-doublet model
assuming that 
the first doublet $\phi_1$
coincides with the Higgs doublet of the SM,
\textit{viz.}\ it does not mix with the second doublet.
Moreover,
we have employed several simplifying assumptions
in order to reduce the parameter space of the model.
We have noted that some finetuning is needed
in order that $\mathrm{BR} \left( \mu^- \to e^- e^+ e^- \right)$
does not become too large when all other four branching ratios
are simultaneously close to their experimental limits.

Flavour-diagonal Yukawa coupling matrices have no straightforward
implementation in the 
quark sector,\footnote{For an attempt in this direction see, however,
  ref.~\cite{GL2003}.}
so one has to admit non-diagonal Yukawa couplings there and 
avoid excessive flavour-changing neutral interactions by finetuning.
Thus there is an asymmetry between the quark and the lepton sector.
This may seem ugly, but, as pointed out in this paper, 
the intriguing consequences for charged-lepton decays make a
consideration of such a framework worthwhile.

\vspace*{5mm}

\paragraph{Acknowledgements:}

E.H.A.\ is supported by the FWF Austrian Science Fund
under the Doctoral Program W1252-N27 ``Particles and Interactions.''
L.L.\ is supported by
the FCT Portuguese Science Foundation
through the projects CERN/FIS-NUC/0010/2015
and UID/FIS/00777/2013,
which are partially funded by POCTI (FEDER),
COMPETE,
QREN,
and the European Union.

\end{document}